\documentclass[a4paper,12pt]{amsart}
\usepackage{amssymb,amsfonts,amscd, amsthm}
\usepackage{graphicx}
\title[Quaternion Landau-Ginsburg models]{Quaternion Landau-Ginsburg models\\
and noncommutative Frobenius manifolds.}
\author[S.M.~Natanzon]{S.M.~Natanzon}

\begin{document}

\newcommand{\glue}{\mathop{\sf glue}\nolimits}
\newcommand{\cut}{\mathop{\sf cut}\nolimits}
\newcommand{\contr}{\mathop{\sf contr}\nolimits}
\newcommand{\Kl}{\mathop{\sf Kl}\nolimits}
\newcommand{\Mb}{\mathop{\sf Mb}\nolimits}
\newcommand{\Cyl}{\mathop{\sf Cyl}\nolimits}
\newcommand{\M}{\mathop{\rm M}\nolimits}
\newcommand{\conv}{\mathop{\sf Contr}\nolimits}
\newcommand{\cor}{\mathop{\rm Cor}\nolimits}
\newcommand{\Hom}{\mathop{\rm Hom}\nolimits}
\newcommand{\Ind}{\mathop{\rm Ind}\nolimits}
\newcommand{\aut}{\mathop{\rm Aut}\nolimits}
\newcommand{\tr}{\mathop{\rm tr}\nolimits}
\newcommand{\ad}{\mathop{\rm ad}\nolimits}
\newcommand{\id}{\mathop{\rm id}\nolimits}
\newcommand{\Ker}{\mathop{\rm Ker}\nolimits}
\newcommand{\Aut}{\mathop{\rm Aut}\nolimits}
\newcommand{\spin}{\mathop{\rm Spin}\nolimits}
\newcommand{\SL}{\mathop{\rm SL}\nolimits}
\newcommand{\GL}{\mathop{\rm GL}\nolimits}
\newcommand{\PSL}{\mathop{\rm PGL}\nolimits}
\newcommand{\PGL}{\mathop{\rm PGL}\nolimits}
\newcommand{\SO}{\mathop{\rm SO}\nolimits}
\newcommand{\Spin}{\mathop{\rm Spin}\nolimits}
\newcommand{\note}{ {\bf Remark.}\quad}

\newtheorem{theorem}{Theorem}[section]
\newtheorem*{definition}{Definition}
\newtheorem{lemma}{Lemma}[section]
\newtheorem{proposition}{Proposition}[section]
\newtheorem{corollary}{Corollary}[section]
\pagestyle{myheadings} \markboth {Quaternion Landau-Ginsburg
models.}
  {Quaternion Landau-Ginsburg models.}
\maketitle

\centerline{Independent University of Moscow, Moscow, Russia}
\centerline{Moscow State University, Moscow, Russia}
\centerline{Institute Theoretical and Experimental Physics, Moscow,
Russia}

\begin{abstract}

We extend the classical topological Landau-Ginsburg model to a
quaternion Landau-Ginsburg model, that satisfy the axioms of
open-closed topological field theory. Later we prove, that a moduli
space of quaternion Landau-Ginsburg models are non-commutative
Frobenius manifold in means of [15].


{\it JGP SC:} Symplectic geometry; Quantum field theory; Strings and
superstrings

\end{abstract}

\vspace{1.8cm}

\section*{1. Introduction}

In this paper we construct the dotted cells and dotted lines in the
next scheme:

\begin{figure}
  \includegraphics[width=9cm]{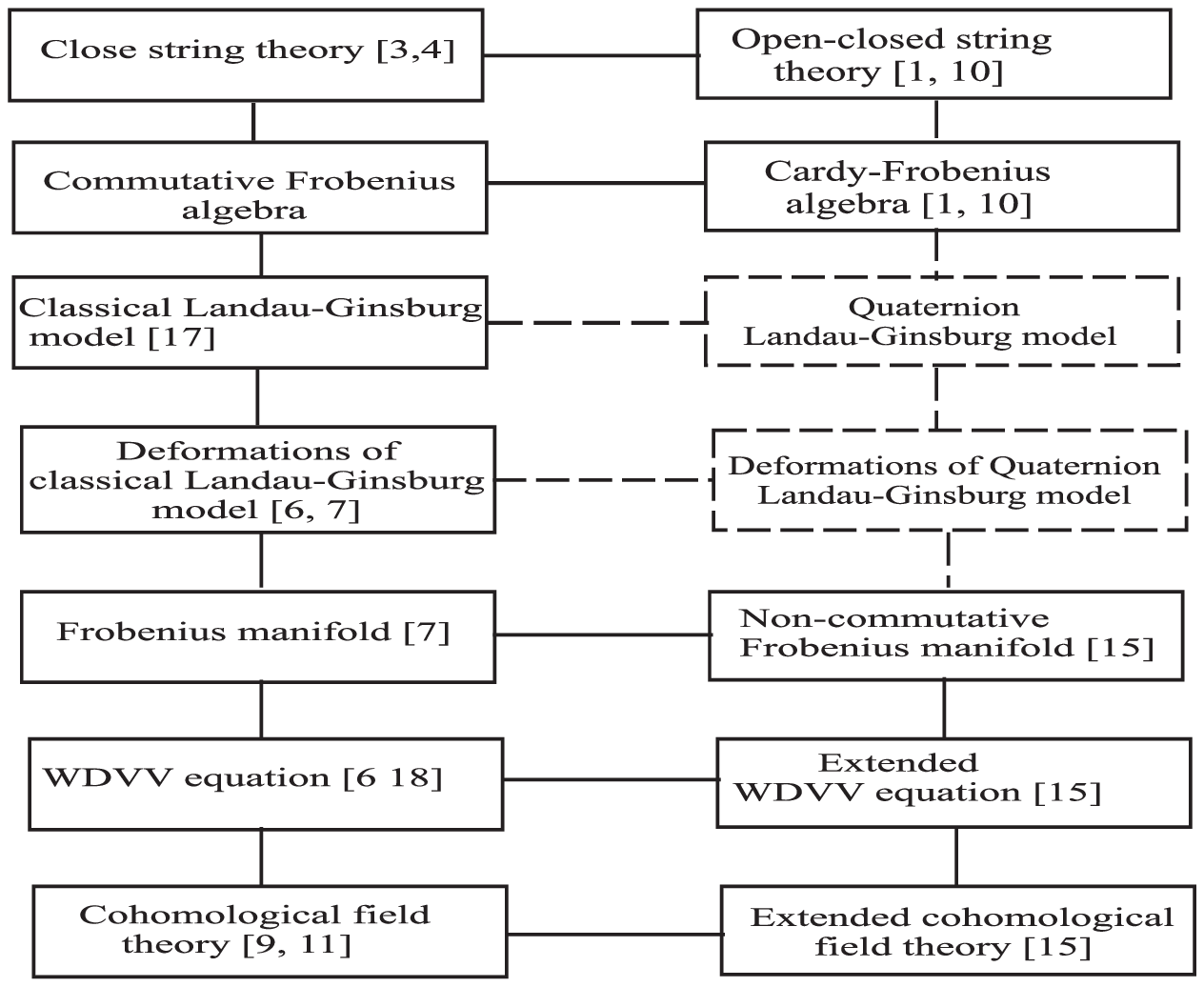}\\
\end{figure}

\vspace{1.8cm}

Explain this scheme in detail. The classical topological
Landau-Ginsburg model was found by Vafa [17](see also Dijkgraaf-
Witten [5]). It generates an  algebra over $\mathbb{C}$ on the
tangent space to a polynomial
$p(z)=z^{n+1}+a_1z^{n-1}+a_2z^{n-2}+...+a_n$ in the space $Pol(n)$
of all such polynomials. This algebra $A_p$ is associative, with the
unity and a linear functional $l_p:A_p\to \mathbb{C}$, such that the
bilinear form $(d_1,d_2)= l_p(d_1d_2)$ is non degenerates. We call
by Frobenius pairs all pairs $(A_p,l_p)$ with such algebraic
properties. Algebra $A_p$ is commutative for classical topological
Landau-Ginsburg model.

Commutative Frobenius pairs one-to-one correspond to topological
field theories that appear from closed topological strings [3, 4, 7,
16]. These topological field theories naturally extend up to
open-closed topological field theories, describing strings with
boundary [10, 12], and even up to Kleinian topological field
theories, describing strings with arbitrary world sheets [1]. In
their turn, open-closed topological field theories one-to-one
correspond to combinations from one commutative and one unrestricted
Frobenius pairs, connected by Cardy condition [1]. We call such
algebraic structure by Cardi-Frobenius algebra. Some classifications
of Cardi-Frobenius algebras is contained in [1].

In the present paper we construct some extension of classical
topological Landau-Ginsburg model to a Cardy-Frobenius algebra with
a quaternion structure. Next we prove, that the set of such
quaternion Landau-Ginsburg models over all polynomials
$p(z)=z^{n+1}+a_1z^{n-1}+ a_2z^{n-2}+...+a_n$ form a non-commutative
Frobenius manifold in means of [15].

Let us explain this result in more detail. The moduli space of the
classical topological Landau-Ginsburg models coincides with the
space $Pol(n)$ of miniversale deformation for the singularity of
type $A_n$ [2]. The metrics $(d_1,d_2)= l_p(d_1d_2)$ on the algebras
$A_p$ turn $Pol(n)$ into a Riemannian manifold with some additional
properties [6, 7]. The differential-geometric structure, arising
here, is an important example of Frobenius manifolds [7, 11]. The
theory of Frobenius manifolds has a lot of applications in different
areas of mathematics (integrable systems, singularity theory,
topology of symplectic manifolds, geometry of moduli spaces of
algebraic curves etc.).

The Dubrovin's theory of Frobenius manifolds [7] is a theory of flat
deformations of commutative Frobenius pairs. As we discussed,
Frobenius pairs are extended up to Cardy-Frobenius algebras. This
suggests on extension of Frobenius manifolds to Cardy-Frobenius
manifolds. An approach to this problem, is contained in [15]. It is
based on Kontsevich-Manin cohomological field theory [9].

In this paper we define Cardy-Frobenius bundles as spaces of some
flat deformations of Cardy-Frobenius algebras and prove, that they
are Cardy-Frobenius (noncommutative) manifolds as regards to [15].
Moreover we prove that the family of all the quaternion
Landau-Ginsburg  models form a Cardy-Frobenius bundle with
quaternion structure.

\section*{2. Cardy-Frobenius algebras.}

Following [10,12] describe the algebraic structure (Cardy-Frobenius
algebra), connected with open-closed topological field theory. It is
follow from [1], that Cardy-Frobenius algebras one-to-one correspond
to open-closed topological field theories.

\subsection*{2.1.Frobenius pairs.}

{\ }

$\textbf{Definition.}$ By {\it Frobenius pair} over a field
$\mathbb{K}$ we call a set $(D,l^D)$ where

1. $D$ is an associative algebra over $\mathbb{K}$ with the unity
element $1^D$.

2. $l^D:D\to \mathbb{K}$ is a $\mathbb{K}$ - linear functional such
that the bilinear form $(d_1,d_2)^D=l^D(d_1d_2)$ is non degenerated.

In this case  $D$ is a Frobenius algebra [8].

$\textbf{Definition.}$ By {\it orthogonal sum} of Frobenius pairs
$(D_1,l^{D_1})$ and $(D_2,l^{D_2})$  we call the Frobenius pair
$(D,l^D)=$ $(D_1,l^{D_1})\oplus (D_2,l^{D_2})$, where $D=D_1\oplus
D_2$ is the direct sum of algebras ($d_1d_2=0$ for $d_i\in D_i$) and
$l^D=l^{D_1}\oplus l^{D_2}$.

$\textbf{Exampele 2.1.}$ $\mathbb{K}$ - numbers.

$\mathbb{K}(\lambda)=(\mathbb{K},l_\lambda)$, where $\lambda\neq
0\in\mathbb{K}$ and $l_\lambda(z)=\lambda z$ for $z\in\mathbb{K}$.

$\textbf{Exampele 2.2.}$ Matrixes over $\mathbb{K}$.

$\mathbb{M}(n,\mathbb{K})(\mu)=(\mathbb{M}(n,\mathbb{K}),l_\mu)$,
where $\mathbb{M}(n,\mathbb{K})$ is the algebra of $n\times n$
$\mathbb{K}$-matrixes, $\mu\neq 0\in\mathbb{K}$ and $l_\mu(z)=\mu
\textmd{tr}(z)$ for $z\in\mathbb{M}(n,\mathbb{K})$.

$\textbf{Exampele 2.3.}$ Quaternions over $\mathbb{K}$.

Let $\mathbb{R}_\mathbb{K}\in\mathbb{K}$ be a subfield, isomorphic
to the field of real numbers. Let $\mathbb{H}_\mathbb{R}$ be the
algebra of quaternions, that is the algebra over $\mathbb{R}$,
generated by vectors $1^\mathbb{H}$, $I$, $J$, $K$, where $IJ=K,
JK=I, KI=J$. Use the isomorphism
$\mathbb{R}_\mathbb{K}\cong\mathbb{R}$ for define
$\mathbb{H}_\mathbb{K}=\mathbb{H}_\mathbb{R}\bigotimes_\mathbb{R}\mathbb{K}$.
We will consider $1^\mathbb{H}$, $I$, $J$, $K$ also as a basis of
$\mathbb{H}_\mathbb{K}$ over $\mathbb{K}$. Put
$\mathbb{H}_\mathbb{K}(\rho)=(\mathbb{H}_\mathbb{K},l_\rho)$ where
$\rho\neq 0\in\mathbb{K}$ and $l_\rho:\mathbb{H}_\mathbb{K}\to
\mathbb{K}$ be the $\mathbb{K}$-linear functional, defined by
$l_\rho(1^\mathbb{H})=2\rho$,
$l_\rho(I)=l^\mathbb{H}(J)=l^\mathbb{H}(K)=0$.

The correspondance
$$\left(%
\begin{array}{cc}
 1 & 0
   \\
 0 & 1 \\
\end{array}%
\right)\mapsto 1^\mathbb{H},
\left(%
\begin{array}{cc}
 -\textit{i} & 0
   \\
 0 & \textit{i} \\
\end{array}%
\right)\mapsto I,
\left(%
\begin{array}{cc}
0 & -1
   \\
 1 & 0 \\
\end{array}%
\right)\mapsto J,
\left(%
\begin{array}{cc} 0 & \textit{i}
   \\
 \textit{i} & 0 \\
\end{array}%
\right)\mapsto K$$ defines an isomorphism between
$\mathbb{M}(2,\mathbb{K})(\rho)$ and $\mathbb{H}_\mathbb{K}(\rho)$.

\subsection*{2.2  Cardy-Frobenius algebras.}

{\ }

$\textbf{Definition.}$  By {\it Cardy-Frobenius algebra} over
$\mathbb{K}$ we call a set $\{(A,l^A),(B,l^B),\phi\}$, where

1. $(A,l^A)$ and $(B,l^B)$ are Frobenius pairs over $\mathbb{K}$ and
$A$ is a commutative algebra;

2. $\phi: A\to B$ is a homomorphism of algebras, such that $\phi(A)$
belong to centre of $B$;

3. For any $x,y\in B$ is fulfilled  $(\phi^*(x),\phi^*(y))^A=
tr(W_{x, y})$, where $\phi^*: B\rightarrow A$ is the linear
operator, such that $(a,\phi^*(b))^A= (\phi(a),b)^B$, for $a\in A,
b\in B$, and $W_{x, y}: B\rightarrow B$ is the linear operator,
defined by $W_{x, y}(b)= xby$.

The condition of this type first appear in works of Cardy and
usually has him name. Let us give a coordinate representation of
Cardy condition, using bases $(\alpha_1,...\alpha_n)\subset A$ and
$(\beta_1,...\beta_n)\subset B$. Let $\{F^{\alpha_i,\alpha_j}\}$ and
$\{F^{\beta_i,\beta_j}\}$ be matrixes inverse to
$F_{\alpha_i,\alpha_j}= (\alpha_i,\alpha_j)^A$ and
$F_{\beta_i,\beta_j}= (\beta_i,\beta_j)^B$. Then condition 3. is
equivalent to

3'. For any $x,y\in B$ is fulfilled
$F^{\alpha_i,\alpha_j}l^B(\phi(a_i)x)l^B(\phi(a_j)y) =
F^{\beta_k,\beta_j} l^B(x\beta_jy\beta_k)= F^{\beta_k,\beta_j}
l^B(\beta_jx\beta_ky)$.

\textsl{Proof of equivalent.} The equality $(\phi^*(x),\phi^*(y))^A=
F^{\alpha_i,\alpha_j}l^B(\phi(a_i)x) l^B(\phi(a_j)y)$ is obviously.
Consider a basis of $B$ such that $(\beta_k,\beta_j)=\delta_{kj}$.
Then $F^{\beta_k,\beta_j}l^B(x\beta_j y\beta_k)=$
$F^{\beta_k,\beta_k}l^B(x\beta_ky\beta_k)= \sum_{k=1}^n (W_{x,
y}(\beta_k),\beta_k)^B= tr(W_{x, y})$.

$\textbf{Exampele 2.4.}$ $\mathbb{K}$ - numbers and matrixes.

$\{\mathbb{K}(\mu^2),\mathbb{M}(n,\mathbb{K})(\mu),\phi_M \}$, where
$\phi_M: \mathbb{K}\to \mathbb{M}(n,\mathbb{K})$  is the naturel
homomorphism of numbers to diagonal matrixes.

\textsl{Check the axiom 3.} Choice the elementary matrixes $E_{ij}$
as a base of $\mathbb{M}(n,\mathbb{K})$. Left and right parts of
axiom 3' are equal to 0, if one of the matrixes  $x$ or $y$ are not
diagonal. Left and right and parts of axiom 3' are equal to $1$ for
$x=E_{ii}$, $y=E_{jj}$.

$\textbf{Definition.}$ By {\it orthogonal sum} of Cardy-Frobenius
algebras $\{(A_1,l^{A_1}),(B_1,l^{B_1}),\phi_1\}$ and
$\{(A_2,l^{A_2}),(B_2,l^{B_2}),\phi_2\}$ we call the Cardy-Frobenius
algebra $\{(A,l^A),(B,l^B),\phi\}=$
$\{(A_1,l^{A_1}),(B_1,l^{B_1}),\phi_1\}$ $\oplus$
$\{(A_2,l^{A_2}),(B_2,l^{B_2}),\phi_2\}$, where $(A,l^A)=$
$(A_1,l^{A_1})\oplus(A_2,l^{A_2})$, $(B,l^B)=$
$(B_1,l^{B_1})\oplus(B_2,l^{B_2})$ and $\phi=\phi_1\oplus\phi_2.$

A Cardy-Frobenius algebra is called {\it semisimple}, if $A$ and $B$
are semisimple algebras. It is follow from [1], that any semisimple
Cardy-Frobenius algebra is isomorphic to orthogonal sum of some
numbers of algebras of
$\{\mathbb{K}(\mu_i^2),\mathbb{M}(n,\mathbb{K})(\mu_i),\phi_M \}$
and some numbers of algebras $\{\mathbb{K}(\lambda_i),0,0\}$.

$\textbf{Exampele 2.5.}$ $\mathbb{K}$ - numbers and quaternions.

$\{\mathbb{K}(\rho^2),
\mathbb{H}_\mathbb{K}(\rho),\phi_\mathbb{H}\}$, where homomorphism
$\phi_\mathbb{H}: \mathbb{K}\to \mathbb{H}$ is defined by
$\phi_\mathbb{H}(1)=1^\mathbb{H}$.

The isomorphism between $\mathbb{M}(2,\mathbb{K})(\rho)$ and
$\mathbb{H}_\mathbb{K}(\rho)$ generate the isomorphism between
$\{\mathbb{K}(\rho^2),\mathbb{M}(2,\mathbb{K})(\rho),\phi_M \}$ and
$\{\mathbb{K}(\rho^2),
\mathbb{H}_\mathbb{K}(\rho),\phi_\mathbb{H}\}$.

\section*{3.  Quaternion Landau-Ginsburg models.}

The classical topological Landau-Ginsburg model [17] of degree $n$
is generated by a complex polynomial $p$ in the form
$p(z)=z^{n+1}+a_1z^{n-1}+a_2z^{n-2}+...+a_n$, such that all roots
$\alpha_1,...,\alpha_n$ of its derivative $p'(z)$ are simple.

The set of all such polynomials form a complex manifold $Pol(n)$ of
complex dimension $n$. Its tangent space $A_p$ in a point $p$
consists of all polynomials of degree $n-1$. The Landau-Ginsburg
model generates on $A_p$ a structure of algebra, where the
multiplication $q=q_1*_pq_2$ for polynomials $q_1=q_1(z)$ and
$q_2=q_2(z)\in A_p$ is defined by condition
$q(z)=q_1(z)q_2(z)(\texttt{mod} p'(z))$.

Moreover, the Landau-Ginsburg model generates on $A_p$ a
non-degenerated bilinear form $(q_1,q_2)_p=l_p(q_1q_2)=
l_p(q_1*_pq_2)$, where $l_p(q)=\frac{1}{2\pi i}\oint\frac {q(z)
dz}{p'(z)}$. (Here and late the formula $\frac{1}{2\pi i}\oint$
means "minus residue in $\infty$"). Thus $A_p$ has a structure of
Frobenius algebra.

The polynomials $e_{p,\alpha_i}(z)=\prod_{j\neq i}
\frac{z-\alpha_j}{\alpha_i-\alpha_j}$ form a basis of idempotents of
$A_p$, that is
$e_{p,\alpha_i}e_{p,\alpha_j}=\delta_{ij}e_{p,\alpha_i}$. Thus $A_p$
is a semi-simple algebra. Put $\mu_{p,\alpha_i}=l_p(e_{p,\alpha_i})=
\frac{1}{n+1}\prod_{j\neq i}\frac{1}{\alpha_i-\alpha_j}$. Let
$A_{p,\alpha_i}$ be the complex vector space generated by
$e_{p,\alpha_i}$ and $l^A_{p,\alpha_i}=l_p|_{A_{p,\alpha_i}}$. Then
$(A_{p,\alpha_i},l^A_{p,\alpha_i})$ is a Frobenius pair, isomorphic
to $\mathbb{C}(\mu_{p,\alpha_i})$.

Consider the Frobenius pair $(B_{p,\alpha_i},l^B_{p,\alpha_i})$ ,
where  $B_{p,\alpha_i}= A_{p,\alpha_i}\otimes_\mathbb{C}
\mathbb{H}_\mathbb{C}\cong\mathbb{H}_\mathbb{C}$ and
$l^B_{p,\alpha_i}$ is defined by
$l^B_{p,\alpha_i}(1^\mathbb{H})=2\rho_{p,\alpha_i}$,
$\rho_{p,\alpha_i}^2=\mu_{p,\alpha_i}$,
$l^B_{p,\alpha_i}(I)=l^B_{p,\alpha_i}(J)=l^B_{p,\alpha_i}(K)=0$. Let
us define the homomorphism
$\phi_{p,\alpha_i}:A_{p,\alpha_i}\rightarrow B_{p,\alpha_i}$ by
$\phi_{p,\alpha_i}(e_{p,\alpha_i})= e_{p,\alpha_i}\otimes_\mathbb{C}
\mathbb{H}_\mathbb{C}$. Then
$\{(A_{p,\alpha_i},l^A_{p,\alpha_i}),(B_{p,\alpha_i},l^B_{p,\alpha_i}),
\phi_{p,\alpha_i}\}$ is a Cardy-Frobenius algebra, isomorphic to
$\{\mathbb{C}(\rho_{p,\alpha_i}^2), \mathbb{H}_\mathbb{C}
(\rho_{p,\alpha_i}),\phi_\mathbb{H}\}$.

Put $B_p= \bigoplus_{i=1}^nB_{p,\alpha_i}=A_p\otimes_\mathbb{C}
\mathbb{H}_\mathbb{C}$. Let $\phi_p: A_p\rightarrow
A_p\otimes_\mathbb{C} \mathbb{H}_\mathbb{C}= B_p$ be the natural
homomorphism. By {\it quaternion Landau-Ginsburg model} we call the
Cardy-Frobenius algebra $\{(A_p,l^A_p),(B_p,l^B_p),
\phi_p\}=\bigoplus_{i=1}^n\{(A_{p,\alpha_i},l^A_{p,\alpha_i}),(B_{p,\alpha_i},
l^B_{p,\alpha_i}), \phi_{p,\alpha_i}\}$.

\section*{4. Frobenius manifolds.}

\subsection*{4.1. Frobenius manifolds and WDVV equations.}

{\ }

In middle of 90 years of the last century B.Dubrovin found and
investigated a class of "flat" deformation of commutative Frobenius
algebras that appear in different domain of mathematics. He calls
this structure Frobenius manifolds [7]. {\it Frobenius manifold} is
a manifold with a Dubrovin connection. Present some equivalent
definitions of the Dubrovin connection.

\textbf{Definition.}([7]) Let $M$ be a smooth (real or complex)
manifold. By {\it Dubrovin connection } on $M$ is called is a family
of commutative Frobenius pairs $(M_p,\theta_p)$ on the tangent
spaces $M_p=T_pM$ for any point $p\in M$ such that

1. Tensors $\theta=\{\theta_p|p\in M\}$, $g(a,b)=\theta(ab)$,
$c(a,b,d)=\theta(abd)$ are smooth and $\textbf{d}\theta=0$.

2. The Levi-Civita connection $\triangledown$ of the matric $g$ is
flat and such that $\triangledown e=0$, where $e$ is the field of
unity elements of the algebras $M_p$.

3. Tensors $c(a,b,d)$ and $\triangledown_fc(a,b,d)$ are symmetrical
by variables $a,b,d,f$.

4. There exists a vector field $E$ (it is called {\it Euler field}),
such that $\triangledown(\triangledown E)=0$.

A Frobenius manifold $M$ is called semi-simple, if $M_p$ is a
semi-simple algebra for any point $p\in M$.

If  $\theta(e)=0$, then Frobenius manifold has a special {\it flat
quasi-homogeneous coordinate system} $t=(t^1,...,t^n)$, such that
$g=\sum_{ij}\delta_{i+j,n+1}\textbf{d}t^i\bigotimes \textbf{d}t^j$,
$e=\partial/\partial t^1$; $d_n=1$, $d_ir_i=0$,
$d_i+d_{n+1-i}=\upsilon+2$ for all $i$; $E=\sum_{i=1}^n
(d_it^i+r_i)(\partial/\partial t^i)$ for some constants $d^i,r_i$.
We call such Dubrovin connection {\it anti-diagonale}.

Recall, that any  semi-simple commutative Frobenius algebra is a
direct sum of one-dimensional one [8]. Thus it has a {\it canonical
basis} $e_1,...,e_n$. This is a basis with the properties
$e_ie_j=\delta_{ij}e_i$, $l(e_i)\neq 0$. The canonical basis is
defined uniquely up to enumeration of its elements.

\textbf{Definition.}([13]) A semi-simple anti-diagonale Dubrovin
connection on a smooth (real or complex) manifold  $M$  is a family
of commutative Frobenius pairs $(M_p,\theta_p)$ on the tangent
spaces $M_p=T_pM$ for any point $p\in M$ such that

1. Tensors $\theta=\{\theta_p|p\in M\}$, $g(a,b)=\theta(ab)$,
$c(a,b,d)=\theta(abd)$ are smooth and $\textbf{d}\theta=0$

2. There exists a covering $M=\bigcup_\alpha U_\alpha$ by coordinate
maps $(x_\alpha^1,...,x_\alpha^n): U_\alpha\rightarrow
\mathbb{K}^n$, such that: a) the vectors $(\partial/\partial
x_\alpha^1,...,\partial/\partial x_\alpha^n)$ form a canonical basis
in $M_p$ for any $p\in U_\alpha$; b) the field $E=\sum_{i=1}^n
x_\alpha^i(\partial/\partial x_\alpha^i)$ don't depend from
$\alpha$; c)$L_E\theta=(\upsilon+1)\theta$, where $L_E$ is the Li
derivative by $E$. Such coordinate system is call  {\it canonical}.

3. The Levi-Civita connection $\triangledown$ of the matric $g$ is
flat. It exists a coordinate system $t=(t^1,...,t^n)$ on $M$, such
that
$g=\sum_{ij}\delta_{i+j,n+1}\textbf{d}t^i\bigotimes\textbf{d}t^j$,
$e=\partial/\partial t^1$ and $E=\sum_{i=1}^n
(d_it^i+r_i)(\partial/\partial t^i)$, where $d_i,r_i$ are constants.

By {\it potential} of Dubrovin connection $(M_p,\theta_p)$ on flat
quasi-homogeneous coordinates $t=(t^1,...,t^n)$, is called a
function $F(t^1,...,t^n)$ such that
$\theta_p(\frac{\partial}{\partial t^i}\frac{\partial}{\partial
t^j}\frac{\partial}{\partial t^k})=\frac{\partial F}{\partial
t^i\partial t^j\partial t^k}$.

\textbf{Definition.} Let $F(t^1,...,t^n)$ be a function on a set
$U\subset\mathbb{C}^n=$ ${(t^1,...,t^n)}$. Let  $E=\sum_{i=1}^n
(d_it^i+r_i)(\partial/\partial t^i)$ be a vector field ,such that
$d_n=1$, $d_ir_i=0$ и $d_i+d_{n+1-i}=\upsilon+2$ for all $i$. Then
the pair $(F,E)$ is a {\it solution of WDVV equations} if:

1.$\sum_{q=1}^n\frac{\partial^3F}{\partial t^i\partial t^j\partial
t^q}$ $\frac{\partial^3F}{\partial t^k\partial t^l\partial
t^{n+1-q}}=$ $\sum_{q=1}^n\frac{\partial^3F}{\partial t^k\partial
t^j\partial t^q}$ $\frac{\partial^3F}{\partial t^i\partial
t^l\partial t^{n+1-q}}$

(associativity equations);

2.$\frac{\partial^3F}{\partial t^i\partial t^j\partial
t^n}=\delta_{i+j,n+1}$

(normalization condition);

3.$L_EF=(\upsilon+3)F+\sum_{ij}A_{ij}t^it^j+\sum_{i}B_it^i+C$, where
$L_E$ is the Li derivative and $A_{ij},B_i,C$ are constants.

(quasi-homogeneous conditions).

This equations was found in works of  Witten [18] and  Dijkgraaf,
E.Verlinde, H.Verlinde [6] for a description of spaces of
deformations of topological fields theories.

According to [7], any solution of  WDVV equations is a potential of
some anti-diagonale Dubrovin connection and moreover the
correspondance (potential)$\mapsto$(anti-diagonale Dubrovin
connection) form one-to-one correspondance between solutions of WDVV
equations and anti-diagonale Dubrovin connections.

If a solution $F$ of WDVV equations has a representation in the form
of Tailor series $F(t)=\sum c(i_1,i_2,...,i_k)
t^{i_1}t^{i_2}...t^{i_k}$, then associativity equations are
equivalent to some relations between the coefficients
$c(i_1,i_2,...,i_k)$. M.Kontsevich, Yu.Manin [9, 11] presented a
family of coefficients with these relations by a special system of
homomorphisms $\mathbb{C}^{\otimes l}\rightarrow
H^*(\overline{M}_{0,l})$, where $M_{0,l}$ is the moduli space of
spheres with $l$ pictures. This gives some other method of
description for Frobenius manifolds. It is called {\it Cohomological
Field Theory}.

\subsection*{4.2. Moduli spaces of classical topological Landau-Ginsburg models.}

{\ }

The first and very important example of a complex Frobenius manifold
is the the moduli space $Pol(n)$ of classical topological
Landau-Ginsburg models. This space appear also in the theory of
singularity, the theory of Coxeter groups, the theory of moduli
spaces of Riemann surfaces, in matrix models of mathematical physics
and in integrable systems. Let us describe this example more
detailed, following on the whole [7].

\textbf{Theorem 4.1.}[6,7]  The structure of Frobenius pairs
$(A_p,l_p)$ for $p\in Pol(n)$ generates a complex Dubrovin
connection on the space $Pol(n)$ of polynomials
$p(z)=z^{n+1}+a_1z^{n-1}+a_2z^{n-2}+...+a_n$ .

Proof. Axiom 1 directly follow from the definitions. Let
$(\alpha_1,...,\alpha_n)$ be the set of roots of $p'$. Consider the
functions $x^{i}(p)=p(\alpha_i)$ as coordinates on $Pol(n)$.

\textbf{Lemma 4.1.} The coordinates $(x^1,...,x^n)$ are canonical,
$\upsilon=\frac{2}{n+1}-1$ and $E=\sum_{i=k}^n\frac{k+1}{n+1}
a_k\partial/\partial a_k$.

Proof. Prove, that $\frac{\partial}{\partial x^j} = e_{p,\alpha_j}$.
Really, $\delta_{ij}= \frac{\partial x^i}{\partial x^j}= $
$\frac{\partial
((\alpha_i)^{n+1}+a_1(\alpha_i)^{n-2}+...+a_n)}{\partial x^j}= $ $
((n+1)(\alpha_i)^n+(n-1)a_1(\alpha_i)^{n-2}+...+a_{n-1})
\frac{\partial\alpha_i}{\partial x^j}+
\frac{\partial\alpha_1}{\partial
x^j}(\alpha_i)^{n-1}+...+\frac{\partial\alpha_n}{\partial x^j}=$
$p'(\alpha_i)\frac{\partial\alpha_i}{\partial x^j}+ \frac{\partial
p}{\partial x^j}(\alpha_i)=\frac{\partial p}{\partial
x^j}(\alpha_i)$. Thus $\frac{\partial p}{\partial
x^j}(\alpha_i)=\delta_{ij}=e_{p,\alpha_j}(\alpha_i)$ and therefore
$\frac{\partial p}{\partial x^j}(z)=e_{p,\alpha_j}(z)$. By
definition, this means that the polynomial $e_{p,\alpha_j}$
correspond to the tangent vector $\frac{\partial}{\partial x^j}$.
Thus $(\partial/\partial x_\alpha^1,...,\partial/\partial
x_\alpha^n)$ form a canonical basis.

Prove now, that $l_p=\frac{\textbf{d}a_1}{n+1}$. Really, considering
the coefficients for $z^{n-1}$ in $\frac{\partial a_1}{\partial
x^i}z^{n-1}+...+\frac{\partial a_n}{\partial x^i} =\frac{\partial
p}{\partial x^i}(z)=e_{p,\alpha_i}(z)$ we find that
$l_p(\partial/\partial
x_\alpha^i)=l_p(e_{p,\alpha_i})=\frac{1}{n+1}\frac{\partial
a_1}{\partial x^i}$. Thus $l_p=\sum_{i=1}^nl_p(\partial/\partial
x_\alpha^i)\textbf{d}x_\alpha^i=\sum_{i=1}^n\frac{1}{n+1}\frac{\partial
a_1}{\partial x^i}\textbf{d}x_\alpha^i=\frac{\textbf{d}a_1}{n+1}$.

Prove, that $E=\sum_{i=1}^n x_\alpha^i(\partial/\partial
x_\alpha^i)$ don't depend from $\alpha$ and
$L_El_p=(\upsilon+1)l_p$. Prove at first, that $L_Ep =
p-\frac{z}{n+1}p'$. Really, the polynomials  $L_Ep $ and $
p-\frac{z}{n+1}p'$ have the same degree $n-1$ and they are the same
values in the points $\alpha_1,...,\alpha_n$, because
$L_E(p)(\alpha_k)=(\sum_{i=1}^n x_\alpha^i(\partial/\partial
x_\alpha^i)(p))(\alpha_k)=x^k=p(\alpha_k)=p(\alpha_k)-\frac{z}{n+1}p'(\alpha_k)$.
Consider now the vector field $F=\sum_{k=1}^n\frac{k+1}{n+1}
a_k\partial/\partial a_k$. Then
$L_F(p)(z)=\sum_{k=1}^n\frac{k+1}{n+1}a_kz^{n-k}=
\sum_{k=1}^n(1-\frac{n-k}{n+1})a_kz^{n-k}=p(z)-\frac{z}{n+1}p'(z)=L_E(p)(z)$.
Thus
$L_E(l_p)=L_F(\frac{\textbf{d}a_1}{n+1})=\frac{1}{n+1}\frac{2}{n+1}\textbf{d}a_1=
\frac{2}{n+1}l_p$.

$\Box$

Construct now a flat coordinate systems  on $Pol(n)$. Consider a
function $\omega=\omega(p,z)$ on $Pol(n)\times \mathbb{C}$ such that
$\omega^{n+1}=p(z)$. Put
$z=\omega+\frac{\tilde{t}^1}{\omega}+\frac{\tilde{t}^2}{\omega^2}+
\frac{\tilde{t}^3}{\omega^3}+...$. The equality
$\omega^{n+1}=p(\omega+\frac{\tilde{t}^1}{\omega}+\frac{\tilde{t}^2}{\omega^2}+
\frac{\tilde{t}^3}{\omega^3}+...)$ gives a possibility to find
$\tilde{t}^i$ as a quasihomogenous  polynomial of $\{a_j\}$. In
particularity $\tilde{t}^1=-\frac{a_1}{n+1}$ и
$\tilde{t}^2=-\frac{a_2}{n+1}$. Put $t^1=-(n+1)\tilde{t}^n$,
$t^n=-\tilde{t}^1$ and $t^i=-\sqrt{n+1}\tilde{t}^{n+1-i}$ for
$i=2,...,n-1$.

\textbf{Lemma 4.2.}  The coordinates $(t^1,...,t^n)$ are  flat
quasi-homogeneous and $d_i=\frac{n+2-i}{n+1}$.

Proof. Consider the set of polynomials
$p(z,\tilde{t})=z^{n+1}+a_1(\tilde{t})z^{n-1}+a_2(\tilde{t})z^{n-2}+...+a_n(\tilde{t})$,
where $\tilde{t}=(\tilde{t}^1,...,\tilde{t}^n)$. Consider the
function
$z(\omega,\tilde{t})=\omega+\frac{\tilde{t}^1}{\omega}+\frac{\tilde{t}^2}{\omega^2}+
\frac{\tilde{t}^3}{\omega^3}+...$, such that
$\omega^{n+1}=p(z(\omega,\tilde{t}),\tilde{t})$. We will consider
$\omega$ and $\{\tilde{t}^1,\dots,\tilde{t}^n\}$ as independent
variables. Then
$0=\frac{\textbf{d}\omega^{n+1}}{\textbf{d}\tilde{t}^i} =
\frac{\partial p}{\partial \tilde{t}^i}+\frac{\partial p}{\partial
z}\frac{\partial z}{\partial \tilde{t}^i} = \frac{\partial
p}{\partial \tilde{t}^i}+p'\frac{1}{\omega^i}$. Thus $\frac{\partial
p}{\partial \tilde{t}^i}=$ $-p'\frac{1}{\omega^i}$. Therefore
$g(\partial/\partial \tilde{t}^i,\partial/\partial \tilde{t}^j)=$
$l_p(\partial/\partial\tilde{t}^i\partial/\partial
\tilde{t}^j)=-\textbf{Res}_{z=\infty}\frac{\frac{\partial
p}{\partial\tilde{t}^i}\frac{\partial p}{\partial
\tilde{t}^j}}{p'}\textbf{d}z=$
$-\textbf{Res}_{z=\infty}\frac{p'\textbf{d}z}{\omega^{i+j}}=$ $
-\textbf{Res}_{\omega=\infty}\frac{\textbf{d}p}{\omega^{i+j}}=$ $
-\textbf{Res}_{\omega=\infty}\frac{\textbf{d}\omega^{n+1}}{\omega^{i+j}}=$
$ (n+1)\delta_{i+j,n+1}$. Thus, $g(\partial/\partial
t^i,\partial/\partial t^j)=\delta_{i+j,n+1}$.

Let $e=\sum_{i=1}^n\rho_i\frac{\partial}{\partial\tilde{t}^i}$ be
the field of unity elements of the algebras $A_p$. Recall, that
$\tilde{t}^1=-\frac{a_1}{n+1}$. Using this fact and lemma 4.1, we
find that $\delta_{\beta,1}=\textbf{d}\tilde{t}^1(\partial/\partial
\tilde{t}^\beta)=$ $-\frac{a_1}{n+1}\textbf{d}a_1(\partial/\partial
\tilde{t}^\beta)=-l_p(\partial/\partial\tilde{t}^\beta)=$
$-g(e,\partial/\partial
\tilde{t}^\beta)=-g(\sum_{i=1}^n\rho_i\frac{\partial}{\partial
\tilde{t}^i},\frac{\partial}{\partial \tilde{t}^\beta})=$
$-\sum_{i=1}^n\rho_ig(\frac{\partial}{\partial
\tilde{t}^i},\frac{\partial}{\partial \tilde{t}^\beta})=$
$-(n+1)\sum_{i=1}^n\rho_i\delta_{i+\beta,n+1}=
-(n+1)\rho_{n+1-\beta}$. Therefore,
$e=-\frac{1}{n+1}\frac{\partial}{\partial \tilde{t}^n}=
\frac{\partial}{\partial t^1}$

It is follow from lemma 4.1. that $L_Ea_i=\frac{i+1}{n+1}a_i$. Thus,
by the definition of $\tilde{t}^i$, we find that
$L_E\tilde{t}^i=\frac{i+1}{n+1}\tilde{t}^i$. Therefore
$L_Et^i=\frac{n+2-i}{n+1}t^i$, and $E=\sum_{i=1}^n
\frac{n+2-i}{n+1}t^i(\partial/\partial t^i)$.

$\Box$

\textbf{Example 4.1.}Find the numbers  $\mu_{p,\alpha_i}$ as some
functions of flat quasi-homogeneous coordinates for $n=2$. If
$p(z)=z^3+a_1z+a_2$, then $t^2=-\tilde{t}^1=\frac{a_1}{3}$. Moreover
$p'(z)=3z^2+a_1$ and
$\alpha_i=\pm\sqrt{-\frac{a_1}{3}}=\pm\sqrt{-t^2}$. Thus,
$\mu_{p,\alpha_i}=\pm\frac{1}{6\sqrt{-t^2}}$.

\textbf{Example 4.2.} The potential of the Frobenius manifold
$Pol(n)$ is a polynomial $F_n$ [7]. Its coefficients was found in
[14]. In particulary
$F_2=\frac{1}{2}(t^1)^2t^2+\frac{1}{24}(t^2)^4$.

\vspace{1.8cm}

\section*{5. Non-commutative Frobenius manifolds.}

\subsection*{5.1. Extended WDVV equations.}

{\ }

A theory of deformations of closed strings is one of sours for the
theory of Frobenius manifolds. Its mathematical equivalent is flat
deformations of commutative Frobenius pairs. But the theory of
closed strings is only part of more general open-closed topological
field theory. It is follow from [1,10,12], that a mathematica
equivalent of a open-closed topological field theory is a
Cardy-Frobenius algebra. A theory of flat deformation for
Cardy-Frobenius algebras was suggested in [15]. It continue the
Kontsevich and Manin approach [9,11] and it gives some extension of
WDVV equations to differential equations on series of
non-commutative variables.

Describe more detail these equations. Let $t=(t^1,...,t^n)$
(respectively $s=(s^1,...,s^m)$ be the standard coordinates on
$A\cong \mathbb{C}^n$ (respectively on $B\cong \mathbb{C}^m$).
Consider the algebras of formal tensor series $F=\sum
c(i_1,i_2,...,i_k|j_1,j_2,...,j_l) t^{i_1}\otimes
t^{i_2}...t^{i_k}\otimes  s^{j_1}\otimes s^{j_2}...s^{j_k}$, где
$c(i_1,i_2,...,i_k|j_1,j_2,...,j_l)\in\mathbb{C}$. Let $F_A$ be the
part of the series $F$ that consists of all monomial without $s^i$ .

Partial derivatives of $F$ are defined by partial derivatives of
monomials.

We consider that $\frac{\partial (t^{i_1}\otimes
t^{i_2}\otimes...\otimes t^{i_k}\otimes  s^{j_1}\otimes
s^{j_2}\otimes...\otimes s^{j_k})}{\partial t^i}$ is the sum of
monomials $t^{i_1}\otimes t^{i_2}\otimes...\otimes
t^{i_{p-1}}\otimes t^{i_{p+1}}...\otimes t^{i_k}\otimes
s^{j_1}\otimes s^{j_2}\otimes ...\otimes s^{j_k})$, such that
$i_p=i$.

Reciprocally $\frac{\partial (t^{i_1}\otimes
t^{i_2}\otimes...\otimes t^{i_k}\otimes  s^{j_1}\otimes
s^{j_2}\otimes...\otimes s^{j_k})}{\partial s^j}$ is the sum of
monomials $t^{i_1}\otimes t^{i_2}\otimes...\otimes t^{i_k}\otimes
s^{j_1}\otimes s^{j_2}\otimes ...\otimes s^{j_{p-1}}\otimes
s^{j_{p+1}}...\otimes s^{j_k}$, such that $j_p=j$.

Put $\frac{\partial^2}{\partial t^i\partial
t^j}=\frac{\partial}{\partial t^i}\frac{\partial}{\partial t^j}$,
$\frac{\partial^2}{\partial t^i\partial
s^j}=\frac{\partial}{\partial t^i}\frac{\partial}{\partial s^j}$,
$\frac{\partial^2}{\partial s^i\partial
s^j}=\frac{\partial}{\partial s^i}\frac{\partial}{\partial s^j}$,
$\frac{\partial^3}{\partial t^i\partial t^j\partial
t^k}=\frac{\partial}{\partial t^i}\frac{\partial}{\partial
t^j}\frac{\partial}{\partial t^k}$.

The definition of the partial derivatives
$\frac{\partial^3(t^{i_1}\otimes\dotsb\otimes t^{i_k}\otimes
s^{j_1}\otimes\dotsb\otimes s^{j_\ell})} {\partial s^i\partial
s^j\partial s^r}$ is more complicated. These partial derivatives are
the sum of monomials $t^{i_1}\otimes\dotsb\otimes t^{i_k} \otimes
s^{k_2}\otimes\dotsb\otimes s^{k_{p-1}}\otimes s^{k_{p+1}}
\otimes\dotsb\otimes s^{k_{q-1}}\otimes s^{k_{q+1}} \otimes
s^{k_\ell}$ such that the sequences
$s^i,s^{k_2},\dotsb,s^{k_{p-1}},s^j,s^{k_{p+1}}, \dotsb,
s^{k_{q-1}},s^r,s^{k_{q+1}},\dotsb,s^{k_\ell}$ and
$(s^{j_1},\dotsb,s^{j_\ell})$ are the same after an cyclic
transposition.

We consider that a monomials $t^{i_1}\otimes\dotsb\otimes
t^{i_k}\otimes s^{j_1}\otimes\dotsb\otimes s^{j_\ell}$ and
$t^{\widetilde i_1}\otimes\dotsb\otimes t^{\widetilde i_k} \otimes
s^{\widetilde j_1}\otimes\dotsb\otimes s^{\widetilde j_\ell}$ are
{\it equivalent}, if $\cup^k_{r=1}i_r=\cup^k_{r=1}\widetilde i_r$
and $\cup^l_{r=1}j_r=\cup^l_{r=1}\widetilde j_r$. Let
$[t^{i_1}\otimes\dotsb\otimes s^{j_\ell}]$ be the equivalent class
of $t^{i_1}\otimes\dotsb \otimes s^{j_\ell}$. The tensor series
$F=\sum c(i_1,...,i_k|j_1,...,j_\ell) a_{i_1}\otimes\dotsb \otimes
b_{j_\ell}$ generate the tensor series $[F]=\sum
c[i_1,...,i_k|j_1,...,j_\ell][a_{i_1}\otimes\dotsb \otimes
b_{j_\ell}]$, where the sum is the sum by all equivalent classes of
monomials and  $c[i_1,...,i_k|j_1,...,j_\ell]$ is the sum of all
coefficients $c(\widetilde i_1,...,\widetilde j_\ell)$,
corresponding monomials from equivalent class
$[a_{i_1}\otimes\dotsb\otimes b_{j_\ell}]$.

We say that a tensor series $F=\sum c(i_1\dotsb i_k|j_1\dotsb
j_\ell) t^{i_1}\otimes\dotsb\otimes t^k\otimes s^{j_1}\otimes
\dotsb\otimes s^{j_\ell}$ satisfy {\it extended WDVV equations} on a
space $H=A\oplus B$, if the following conditions hold

1. The coefficients $c(i_1\dotsb i_k|j_1\dotsb j_\ell)$ are
invariant under all permutation of $\{i_r\}$.

2. The coefficients $c(i,j|)$ и $c(|i,j)$ generate nondegenerate
matrices. By $F_a^{t^i t^j}$ и $F_b^{s^i s^j}$ denote the inverse
matrices of $c(i,j|)$ and $c(|i,j)$ respectively.

3. $$[\sum_{p,q=1}^n\frac{\partial^3 F_A} {\partial t^i \partial t^j
\partial t^p}\otimes F_a^{t^p t^q} \frac{\partial^3 F_A}{\partial
t^q\partial t^k
\partial t^\ell}]=[\sum_{p,q=1}^n\frac{\partial^3
F_A}{\partial t^k\partial t^j \partial t^p} \otimes F_a^{t^p
t^q}\frac{\partial^3 F_A}{\partial t^q\partial t^i
\partial t^\ell}].$$

4. $$[\sum_{p,q=1}^m\frac{\partial^3 F}{\partial s^i
\partial s^j
\partial s^p}\otimes F_b^{s^p s^q}\frac{\partial^3 F}{\partial s^q\partial s^k
\partial s^\ell}]=
\sum_{p,q=1}^m[\frac{\partial^3 F}{\partial s^\ell\partial s^i
\partial s^p} \otimes F_b^{s^p s^q}\frac{\partial^3
F}{\partial s^q\partial s^j
\partial s^k}].$$

5. $$[\sum\frac{\partial^2 F}{\partial t^k
\partial s^p} \otimes F_b^{s^p s^q}\frac{\partial^3
F}{\partial s^q\partial s^i \partial s^j}]= [\sum\frac{\partial^2
F}{\partial t^k\partial s^p}\otimes F_b^{s^p s^q} \frac{\partial^3
F}{\partial s^q\partial s^j \partial s^i}].$$

6. $$[\sum\frac{\partial^2 F}{\partial s^k
\partial t^p} \otimes F_a^{t^p t^q}\frac{\partial^3
F}{\partial t^q\partial t^i \partial t^j}]= [\sum\frac{\partial^2
F}{\partial t^i\partial s^p}\otimes F_b^{s^p s^q} \frac{\partial^3
F}{\partial s^q\partial s^k
\partial s^r}\otimes F_b^{s^r s^\ell}\frac{\partial^2
F}{\partial s^\ell\partial t^j}].$$

7. $$[\sum\frac{\partial^2 F}{\partial s^u
\partial t^p} \otimes F_a^{t^p t^q}\frac{\partial^2 F}{\partial t^q\partial
s^v}]= [\sum\frac{\partial^3 F}{\partial s^u\partial s^p\partial
s^r}F_b^{s^r s^l}\otimes F_b^{s^p s^q} \frac{\partial^3 F}{\partial
s^l\partial s^v\partial s^q}].$$

In [15] are demonstrated that solutions of extended WDVV equations
one-to-one correspond to potentials of some extension of
Cohomological Field Theory and moreover they describe some class of
deformations of Cardy-Frobenius algebras. Thus it is natural to
consider solutions of extended WDVV equations as (non commutative)
extension of Frobenius manifolds, that we call {\it Cardy-Frobenius
manifolds}. Late we prove that quaternion Landau-Ginsburg models
generate a Cardy-Frobenius manifold.

\subsection*{5.2. Cardy-Frobenius bundles.}

{\ }

\textbf{Definition.} Let $M$ be a , were $\mathbb{K}$ is the real or
the complex field. By {\it Cardy-Frobenius bundle} on smooth (real
or complex) manifold $M$ we call a pair of bundles $\varphi_A: A\to
M$ and $\varphi_B:B\to M$ with a flat connection $\nabla_B$, and a
set of Cardy-Frobenius algebras
$\{(A_p,l^A_p),(B_p,l^B_p),\phi_p\}$, where $A_p=\varphi_A^{-1}(p)$,
$B_p=\varphi_B^{-1}(p)$, such that:

1. The algebras $\{(A_p,l^A_p)\}$ form a Dubrovin connection.

2. The connection $\nabla_B$ conserve the family of bilinear forms
$\{(b_1,b_2)_p=l^B_p(b_1b_2)|p\in M\}$.

Call by {\it a flat system of coordinates} on $B$ a family of linear
coordinates systems $s=\{s_p=(s^1_p,...,s^m_p)|p\in M\}$ on bands
$B_p$, that is invariant by $\nabla_B$. It generates a basis
$(\frac{\partial}{\partial s^1_p},...,\frac{\partial}{\partial
s^m_p})$ on any vector space $B_p$. Axiom 2. say that values
$(\frac{\partial}{\partial s^i_p},\frac{\partial}{\partial
s^j_p})_p=l^B_p(\frac{\partial}{\partial
s^i_p}\frac{\partial}{\partial s^j_p})$ are constants on $M$.

3. Let $s=(s^1,...,s^m)$ be a flat system of coordinates on $B$.
Then the tensor fields $c^B(\frac{\partial}{\partial
s^i},\frac{\partial}{\partial s^j},\frac{\partial}{\partial
s^k})=l^B_p(\frac{\partial}{\partial s^i}\frac{\partial}{\partial
s^j}\frac{\partial}{\partial s^k})$, are smooth as functions on $M$.
We call {\it $B$-structures tensors}.

4. The natural map $\phi=\{\bigcup \phi_p|p\in M\}:A\to B$ is
smooth. It define smooth {\it transition tensors field}
$c^{AB}(a,b)=l^B_p(\phi(a)b)$.

\textbf{Theorem 5.1.} Let $M$ be a semi-simple Frobenius manifold
with a Dubrovin connection  $\{(A_p,l^A_p)|p\in M\}$. Then there
exist Cardy-Frobenius bundles $\{(A_p,l^A_p),(B_p,l^B_p),\phi\}$.

Proof. Let $(x_\alpha^1,...,x_\alpha^n)$ be canonical coordinates on
$M$. Put $\lambda_{p,i}=l^A_p(\partial/\partial x_\alpha^i(p))$. Let
$\mu_{p,i}$ be a smooth function on $M$, such that
$\mu_{p,i}^2=\lambda_{p,i}$. Consider the family of Frobenius pair
$(B_{p,i},l_{p,i}^B)=\mathbb{M}(m,\mathbb{C})(\mu_{p,i})$ from
example 2.2. Put $(B_{p},l_{p}^B)=\bigoplus_i(B_{p,i},l_{p,i}^B)$
and $B=\bigoplus_{p\in M}B_p$. Describe a connection $\nabla_B$. The
standard basis  $\{E^{kr}\}$ of $\mathbb{M}(m,\mathbb{C})$ generate
the basis the $\{E_{p,i}^{kr}\}$ of $B_{p,i}$. We will be consider,
that the connection $\nabla_B$ generate the transfer $E_{p,i}^{kr}$
to $\frac{\mu_{p,i}}{\mu_{q,i}}E_{q,i}^{kr}$ for $q$ from a
neighborhood of $p$. Define a structure of smooth manifold on $B$
considering that the projection $\varphi_B(B_p)=p$ is smooth. Define
the homomorphism $\phi_p: A_p\to B_p$, considering that
$\phi_p(\partial/\partial x_\alpha^i)$ is the unit element of the
algebra $B_{p,i}$. According to example 2.4. the structure, that we
constructed, is Cardy-Frobenius bundles.

$\Box$

\textbf{Definition.} Let $\varphi_A: A\to M$, $\varphi_B:B\to M$,
$\nabla_B$, $\{(A_p,l^A_p), (B_p,l^B_p),\phi_p|p\in M\}$ be a
Cardy-Frobenius bundle on $M$. Let $t=(t^1,...,t^n)$ be a system of
flat quasi-homogenies coordinates of the Dubrovin connection
$(A_p,l^A_p)$ and let $s=(s^1,...,s^m)$ be a system of flat
coordinates on $B$. By {\it potential} of this Cardy-Frobenius
bundle is called the formal tensor series $F(t|s)=\sum
c(i_1,i_2,...,i_k|j_1,j_2,...,j_l) t^{i_1}\otimes
t^{i_2}...t^{i_k}\otimes  s^{j_1}\otimes s^{j_2}...s^{j_k}$, where
$c(i_1,i_2,...,i_k|j_1,j_2,...,j_l)\in\mathbb{C}$, such that

1. The matrixes $c(i,j|)=l^A_p(\frac{\partial}{\partial
t^i}\frac{\partial}{\partial t^j})$ и
$c(|i,j)=l^B_p(\frac{\partial}{\partial s^i}\frac{\partial}{\partial
s^j})$ are non-degenerated. Let $F_a^{t^i t^j}$ and $F_b^{s^i s^j}$
be the matrixes inverted to $c(i,j|)$ and $c(|i,j)$ respectively.

2. If $F_A$ is the part of $F$ that don't depend from $s$, that it
pass to the potential of Dubrovin connection $(M_p,l^A_p)$ after
changing the tensor multiplication to the ordinary multiplication.

3. The formal tensor series $\frac{\partial^3F} {\partial
s^i\partial s^j\partial s^r}$ are not depend from $s$ and, after
changing the tensor multiplication to the ordinary multiplication,
coincide with the $B$-structure tensors of the bundle.

4. The formal tensor series $\frac{\partial^2F} {\partial
t^i\partial s^j}$ are not depend from $s$ and , after changing the
tensor multiplication to the ordinary multiplication, coincide with
the transition function of the bundle.

\textbf{Theorem 5.2.} Let a Dubrovin connection $(A_p,l^A_p)$ on $M$
has a potential in form of Taylor series. Then any Cardy-Frobenius
bundle $\varphi_A: A\to M$, $\varphi_B:B\to M$, $\nabla_B$, $\{(A_p
,l^A_p), (B_p,l^B_p),\phi_p|p\in M\}$ also has a potential.

Proof. Let $t=(t^1,...,t^n)$ be a flat quasi-homogeneous coordinates
of the Dubrovin connection $\{(A_p ,l^A_p)|p\in M\}$. Let
$s=(s^1,...,s^m)$ be be a flat coordinates system on $B$. Consider
the potential of the Dubrovin connection $\{(A_p ,l^A_p)|p\in M\}$.
Changing the ordinary multiplication to the tensor multiplication we
obtain a tensor series $F_A$. Let $F_A^i$ be the tensor series, such
that $\frac{\partial F_A^i} {\partial t^i}=F_A$. Then  the tensor
series $F=F_A+\frac{1}{2}\sum l^A(\frac{\partial}{\partial
t^i}\frac{\partial}{\partial t^j})t^i\otimes t^j+$ $\sum
l^B(\frac{\partial}{\partial s^i}\frac{\partial}{\partial
s^j})s^i\otimes s^j+\sum F_A^it^i\otimes s^i +$ $\frac{1}{3}\sum
l^B(\frac{\partial}{\partial s^i}\frac{\partial}{\partial
s^j}\frac{\partial}{\partial s^k})s^i\otimes s^j\otimes s^k$ is the
potential of the Cardy-Frobenius bundle.

$\Box$

\textbf{Theorem 5.3.} The potential of any  Cardy-Frobenius bundle
satisfy the extended WDVV equations.

Proof. All relation follow from properties of Cardy-Frobenius
algebras $\{(A_p l^A_p),(B_p,l^B_p),\phi_p\}$. In particulary
relation 1 follow from the commutativity of $A$. Relation 2 follow
from the non-degeneracy the bilinear forms on $A$ and $B$. Relations
3 and 4 follow from the associativity of algebras $A$ and $B$.
Relation 5. is true because $\phi_p(A_p)$ belong to center of
algebra $B_p$. Relation  6 is true because the map $\phi_p$ is a
homomorphism. Relation  7 follow from Cardy axiom.

$\Box$

\subsection*{5.3. Moduli space of quaternion Landau-Ginsburg models.}

{\ }

Let us use the construction from theorem 5.1. for quaternion
Landau-Ginsburg models. Then we have

\textbf{Corollary 5.1.} Family of quaternion Landau-Ginsburg models
$\{(A_p,l^A_p), (B_p,l^B_p,),\phi_p|p\in Pol(n)\}$ form a
Cardy-Frobenius bundles over $Pol(n)$. The space $B$ has a natural
quaternion structure that is invariant by the connection $\nabla_B$.

A flat system of coordinates $t=(t^1,...,t^n)$ for $A$ is ,
described in section 4. A basis on $B_p$ is
$\{1^\mathbb{H}e_{p,\alpha_i}, Ie_{p,\alpha_i},Je_{p,\alpha_i},
Ke_{p,\alpha_i}|i=1,...,n\}$ from section 3. In a neighborhood of
$p$ it generate a flat coordinate system
$s_{q,i}=(s_{q,i}^{1^\mathbb{H}},s_{q,i}^I,s_{q,i}^J,s_{q,i}^K)=
\frac{\rho_{p,\alpha_i}}{\rho_{q,\alpha_i}}(1^\mathbb{H}e_{p,\alpha_i},
Ie_{p,\alpha_i},Je_{p,\alpha_i})$, where $\rho_{p,\alpha_i}^2=
\mu_{p,\alpha_i}$.

\textbf{Example 5.1.} For any $V,W\in\{1^{\mathbb{H}},I,J,K\}$,
$\frac{\partial}{\partial s_{h,i}^V} \frac{\partial}{\partial
s_{q,i}^W}=\delta_{h,q}\delta_{i,j} $ $
\frac{\rho_{p,\alpha_i}}{\rho_{q,\alpha_i}} \frac{\partial}{\partial
,s_{q,i}^{VW}}$

Thus we can to find the bilinear form  $(\frac{\partial}{\partial
s^{ij}},\frac{\partial}{\partial
s^{kl}})^B=l^B(\frac{\partial}{\partial
s^{ij}}\frac{\partial}{\partial s^{kl}})$ and the structure tensor
$c^B(\frac{\partial}{\partial s^{ij}},\frac{\partial}{\partial
s^{kl}},\frac{\partial}{\partial s^{uv}})=$
$l^B(\frac{\partial}{\partial s^{ij}}\frac{\partial}{\partial
s^{kl}}\frac{\partial}{\partial s^{uv}})$, by values
$\rho_{q,\alpha_i}$. Example 4.1. contain an algorithm for these
calculations for $n=2$. In this case $\rho_{q,\alpha_i}^2=
\pm\frac{1}{6\sqrt{-t^2}}$.

\textbf{Example 5.2.} Coupling between canonical $x=(x^1,...,x^n)$
and flat quasi-homogeneous coordinates $t=(t^1,...,t^n)$ generate
the transition tensors. Demonstrate this for $n=2$.

According to our definitions, $\frac{\partial}{\partial
t^1}=\frac{\partial}{\partial x^1}+$ $\frac{\partial}{\partial x^2}$
и $\frac{\partial}{\partial t^2}=R_1\frac{\partial}{\partial x^1}+$
$R_2\frac{\partial}{\partial x^2}$. Thus $\frac{\partial}{\partial
t^2}\frac{\partial}{\partial t^2}=$ $R_1^2\frac{\partial}{\partial
x^1}+ R_2^2\frac{\partial}{\partial x^2}$. On the other hand,
according to example 4.2., $\frac{\partial}{\partial
t^2}\frac{\partial}{\partial t^2}=$ $t^2\frac{\partial}{\partial
t^1}=t^2(\frac{\partial}{\partial x^1}+$ $\frac{\partial}{\partial
x^2})$. Thus $R_i=\pm\sqrt{t^2}$.

It is follow from example 4.2., that
$F_A=\frac{1}{2}(t^1)^2t^2+\frac{1}{24}(t^2)^4$. Therefore
$F_A^1=\frac{1}{6}(t^1)^3t^2+\frac{1}{24}t^1(t^2)^4$ and
$F_A^2=\frac{1}{4}(t^1)^2(t^2)^2+\frac{1}{120}(t^2)^5$.

Thus, we can to find the potential of the Cardy-Frobenius bundle for
$n=2$. According to theorem 5.2 it is the tensor series

$F=F_A+\frac{1}{2}\sum l^A(\frac{\partial}{\partial
t^i}\frac{\partial}{\partial t^j})t^i\otimes t^j+$ $\sum
l^B(\frac{\partial}{\partial s^i}\frac{\partial}{\partial
s^j})s^i\otimes s^j+\sum F_A^it^i\otimes s^i +$ $\frac{1}{3}\sum
l^B(\frac{\partial}{\partial s^i}\frac{\partial}{\partial
s^j}\frac{\partial}{\partial s^k})s^i\otimes s^j\otimes s^k$.

{\ }

\textbf{Acknowledgements.} Part of this paper was written during the
author stay at MPIM in Bonn and IHES in Bures-sur-Yvette. I thank
these organisations for support and hospitality. I would like to
thank M. Kontsevich, S. Novikov and A. Schwarz  for useful
discussions of results. This research is partially supported by
grant RFBR-04-01-00762, NSh-1972.2003.1 and NWO 047.011.2004.026.

{\ }

\textbf{References}

1. A.Alekseevskii, S.Natanzon, Noncommutative two-dimensional
topological field theories and Hurwitz numbers for real algebraic
curves. http://xxx.lanl.org/math.GT/0202164.

2. V.Arnold, S. Gusein-Zade, A.Varchenko. Singularities of
differentiable maps, vols I,II. Birkhauser,Boston,1985 and 1988.

3. M.Atiyah, Topological quantum field theories, Inst.Hautes Etudes
Sci. Publ. Math., 68(1988), 175-186.

4. R.Dijkgraaf, Geometrical approach to two-dimensional conformal
field theory, Ph.D. Thesis (Utrecht, 1989).

5. R.Dijkgraaf, E.Witten,  Mean field theory theory, topological
field theory theory and multi-matrix models, Nucl.Phys., B342
(1990), 486.

6. R.Dijkgraaf, E.Verlinde, H.Verlinde, Note on topological string
theory and 2D quantum gravity, Nucl. Phys.B352(1991) 59.

7. B.Dubrovin, Geometry of 2D topological field theories In: LNM,
1620 (1996), 120-348.

8. C.Faith, Algebra: rings, modules and categories I.
Springer-Verlag, Berlin Heidelberg New York, 1973.

9. M.Kontsevich, Yu.Manin, Gromov-Witten classes, quantum
cohomology, and enumerative geometry, Commun. Math. Phys, 164
(1994), 525-562.

10. C.I.Lazorau, On the structure of open-closed topological field
theory in two-dimension Nucl.Phys. B603(2001),497-530.

11. Yu.Manin, Frobenius Manifolds, Quantum Cohomology, and Moduli
Spaces, vol.47, American Mathematical Society, Colloquium
Publication, 1999.

12. G.Moor, D-branes, RR-Fields and K-theory.
http://online.itp.ucsb.edu/online/mp01/moore2/

13. S.Natanzon, Structures de Dubrovin. Preprint 1997/26 IRMA
Strasbourg, 1997.

14. S.Natanzon, Formulas for  $A_n$ and $B_n$ - solutions of WDVV
equations,J. Geom. Phys, 39 (2001), 323-336.

15. S.Natanzon,  Extended cohomological field theories and
noncommutative Frobenius manifolds, J. Geom. Phys, 51 (2003),
387-403.

16. V. Turaev, Homotopy field theory in dimension 2 and
group-algebras. http://xxx.lanl.org/math.GT/9910010.

17. C.Vafa, Topological Landau-Ginsburg models. Modern Physics
Letters A, Vol 6, N 4 (1991), 337-349.

18. E.Witten, On the structure of the topological phase of
two-dimensional gravity, Nucl.Phys., B340 (1990), 281

\

\ natanzon@mccme.ru
\end{document}